# Boost Distribution System Restoration with Emergency Communication Vehicles Considering Cyber-Physical Interdependence

Zhigang Ye, *Member, IEEE*, Chen Chen, *Senior Member, IEEE*, Ruihuan Liu, *Student Member, IEEE*, Kai Wu, *Senior Member, IEEE*, Zhaohong Bie, *Senior Member, IEEE*, Guannan Lou, *Member, IEEE*, Wei Gu, *Senior Member, IEEE*, and Yubo Yuan, *Member, IEEE*

*Abstract*—Enhancing restoration capabilities of distribution systems is one of the main strategies for resilient power systems to cope with extreme events. However, most of the existing studies assume the communication infrastructures are intact for distribution automation, which is unrealistic. Motivated by the applications of the emergency communication vehicles (ECVs) in quickly setting up wireless communication networks after disasters, in this paper, we propose an integrated distribution system restoration (DSR) framework and optimization models, which can coordinate the repair crews, the distribution system (physical sectors), and the emergency communication (cyber sectors) to pick up unserved power loads as quickly as possible. Case studies validated the effectiveness of the proposed models and proved the benefit of considering ECVs and cyber-physical interdependencies in DSR.

## Nomenclature

**Sets**

| | |
|---|---|
| $\mathcal{D}^C, \mathcal{D}^R$ | Depots with emergency communication vehicles (ECVs)/repair crews (RCs) |
| $\mathcal{W}^C, \mathcal{W}^R$ | Working sites (WSs) for ECVs/RCs, where $\mathcal{W}^R$ include switches ($\mathcal{SW}$) and faulted lines ($\mathcal{F}$), i.e. $\mathcal{W}^R = \mathcal{SW} \cup \mathcal{F}$ |
| $\mathcal{W}^C_{(i,j)'}$ | Working sites that can cover the feeder terminal unit (FTU) $(i,j)'$ |
| $\mathcal{N}^C$ | Communication nodes |
| $\mathcal{N}^C_w$: | Communication nodes covered by the ECV at WS $w$ |
| $\mathcal{SW}$ | Switches, including automatic switches ($\mathcal{AS}$) and manual switches ($\mathcal{MS}$), i.e. $\mathcal{SW} = \mathcal{AS} \cup \mathcal{MS}$ |
| $\mathcal{C}^E, \mathcal{C}^L$ | All the electrical node cells, and node cells with loads |
| $\mathcal{L}_c$ | Loads in the node cell $c$ |

**Parameters**

| | |
|---|---|
| $n^{CA}, n^{RA}$ | Dimension of the route table of communication agents (CAs)/repair agents (RAs) |
| $T_0$ | Start time for ECVs and RCs departing from depots |
| $T^{MAX}$ | The scheduled time horizon |
| $T^C_{ij}, T^R_{ij}$ | Travel time for ECVs/RCs traveling from $i$ to $j$ |
| $T^{Cmin}_j$ | Minimum duration of stay for ECVs at the working site $j$ |
| $T^{RP}_f$ | Repair time for RCs to repair the faulted component $f$ |
| $T^{MS}_{(i,j)}$ | Operation time for RCs to manually operate switch $(i,j)$ |
| $T^{AS}_{(i,j)}$ | Operation time of operating switch $(i,j)$ remotely |

**Variables**

| | |
|---|---|
| $x^C_{ij}, x^R_{ij}$ | Elements of the route table of CAs/RAs |
| $t^{Ca}_i, t^{Cd}_i$ | Arrival/departure time of CAs at/from the WS $i$ |
| $t^R_i$ | Arrival time of RAs at the WS $i$ |
| $f^R_i$ | Repair completion time of the node cell $i$ |
| $t^E_i$ | The time when the node cell $i$ is energized) |
| $d^{AO}_{ij}, d^{MO}_{ij}$ | Binaries indicating if the switch $(i,j)$ is automatically/manually operated from $i$ to $j$. |
| $z^C_{(i,j)',k}$ | Binary indicating if FTU $(i,j)'$ is controlled via the ECV at WS $k$ |
| $t^{AOop}_{ij}$ | The time when the automatic switch $(i,j)$ is remotely operated from $i$ to $j$ |
| $d^{AOe}_{ij}, d^{AOde}_{ij}$ | Binaries indicating if the switch $(i,j)$ is automatically operated from $i$ to $j$ with/without electricity |
| $d^{MOe}_{ij}, d^{MOde}_{ij}$ | Binaries indicating if the switch $(i,j)$ is manually operated from $i$ to $j$ with/without electricity |

## I. Introduction

LARGE scale power outages caused by extreme events, such as natural disasters, cascading failures, and cyber-attacks, are becoming more and more frequent throughout the world, which reflects the deficiency of power systems to cope with extreme events of high impact low probability (HILP) [1-3]. Thus, the ways to improve resilience, including preparedness, resistance, responsiveness, and rapid restoration capabilities of

Manuscript received xxx; revised xxx; accepted xxx. Date of publication xxx; date of current version xxx. The work was supported in part by the Science and Technology Project of the State Grid Corporation of China under Grant 5400-202199523A-0-5-ZN; in part by the Outstanding Postdoctoral Program of Jiangsu Province under Grant 2022ZB795; and in part by the National Science Foundation of Jiangsu Province under Grant BK20220217. Paper no. TSG-01843-2021. *(Corresponding author: Chen Chen)*

Zhigang Ye is with the State Grid Jiangsu Electric Power Company Ltd. Research Institute, Nanjing 211103, China; and with the School of Electrical Engineering, Xi'an Jiaotong University, Xi'an 710049, China; and with the School of Electrical Engineering, Southeast University, Nanjing 210096, China (e-mail: yzhggoodluck@hotmail.com).

Chen Chen, Ruihuan Liu, Kai Wu and Zhaohong Bie are with the School of Electrical Engineering, Xi'an Jiaotong University, Xi'an 710049, China (e-mail: morningchen@xjtu.edu.cn; rhliu@stu.xjtu.edu.cn, wukai@mail.xjtu.edu.cn; zhbie@mail.xjtu).

Guannan Lou and Wei Gu are with the School of Electrical Engineering, Southeast University, Nanjing 210096, China (e-mail: bingzhi0828@163.com; wgu@seu.edu.cn).

Yubo Yuan is with the State Grid Jiangsu Electric Power Company Ltd. Research Institute, Nanjing 211103, China (yyb97104@sina.com).

power systems, have been causing increasing attention. Compared to transmission systems, distribution systems are more vulnerable to disasters and closer to customers, leading to most of the power outages occurring in distribution systems [4]. Hence, enhancing the restoration capability of distribution systems is key to improving the resilience of the whole power system [5].

With the development of smart grid technology, the effective and efficient interaction of power network infrastructures (physical systems) with information sensing, processing, intelligence, and control systems (cyber systems) features modern distribution systems [6]. The increased situational awareness with smart meters, phasor measurement units (PMUs), and advanced control capabilities such as automatic feeder switching, have begun to increase the resilience of distribution systems through more effective and timely fault detection, isolation, and restoration (FDIR) [3, 7].

Distribution system restoration (DSR) aims to restore power service through energization paths from power sources to unserved customers [4]. Normally, the cyber sectors of distribution automation (DA) are intact so that quick DSR can be realized through network topology reconfiguration by remotely monitoring and controlling the automatic switches via feeder terminal units (FTUs). However, in the case of extreme events such as natural disasters, both the cyber and physical sectors of the integrated cyber-physical distribution power systems (CPDS) might be disrupted [8, 9]. In such cases, quick service restoration through reconfiguration may not be realized because: 1) the disrupted communication network disables the DA to fulfill the distribution system's self-healing capabilities; 2) an energization path may not exist because multiple physical damages may block all the possible energization paths. Thus, to pick up the unserved customers, the repair crews should go to the damaged components or switches to repair or operate them manually. In recent years, there have been extensive studies on the co-optimization and coordination of DSR and crew dispatch after large-scale power outages, represented by Arif [10], Chen [11], Lei [12], et al. However, all of these work assume that the cyber sectors of the CPDS are intact so that the automatic switches can be controlled remotely, which is too ideal to be applied in the real-world DSR.

The wireless communication network has been widely adopted in the existing distribution automation (DA) systems to support the communication and control between the control center and controllable devices (such as remotely controlled switches) via FTUs. After an extreme event, the wireless communication infrastructure, e.g., base stations, may damage. In such cases, the self-healing capability of the distribution system via DA will be ineffective due to disconnected communication links between FTUs and the control center, leading to the switches failing to be controlled remotely. This may further prolong the restoration process of the power grid so that the intact FTUs, which are originally powered by the electric grid to be restored, will deplete their backup batteries. On the contrary, if we can restore the communication networks before the depletion of the backup batteries of FTUs, the functionality of DA would be enabled for service restoration.

Compared with abundant research basis of power system resilience, there are few studies on the resilience of cyber-physical power systems (CPPS). From the perspective of information systems, [13] proposed a CPPS robust routing model with a priority mechanism that considers cyber-physical disturbances based on robust optimization. [14] proposed a self-healing phasor measurement unit (PMU) network that exploits the programmable configuration in a software-defined networked infrastructure to achieve resilience. [15] developed a model to find the optimal routing in the communication network to minimize the impact of cascading effects triggered by initial failures. From the perspective of the physical power grid, [16] proposed a cooperative evolutionary algorithm that simultaneously evolves a population of unmanned aerial vehicles (UAV) scheduling solutions and a population of human team scheduling solutions. The power–communication coordination recovery strategies based on the gridding method after disasters are proposed in [7]. [17] established an integer linear programming model for DC optimal power flow considering the information network constraints and a multi-stage bi-level model for cyber-physical collaborative recovery. [18] proposed a cyber-constrained optimal power flow model for the emergency response of smart grids. These studies consider the coupling models of the cyber layer and the physical layer but do not fully consider the coupling relationship between the two layers during the power system restoration process.

The emergency communication services, such as the wireless communication networks set up by emergency communication vehicles (ECVs) or UAVs, have been successfully utilized after natural disasters in certain areas where communication restoration is urgent and repairing communication infrastructures takes substantial time and effort [19]. Similarly, emergency communication techniques with ECVs have the potential to enhance the DSR by setting up self-healing communication networks for DA. However, how to utilize emergency communication to coordinate with the service restoration and crew dispatch has not been well considered in the literature.

Hence, motivated by the abovementioned challenges, in this paper, we formulate the interdependencies and cooperation models of emergency communication set-up with ECVs and distribution system restoration. We aim to find extensive ways to enhance the resilience of power systems by considering the effective and efficient integration and interaction of cyber-physical power systems.

The main contributions of this paper can be summarized below:

1) We propose an integrated framework in which setting up emergency communication networks by ECVs, dispatching repair crews, and operating switches in the distribution systems are coordinated to enhance the restoration capabilities of distribution systems.

2) We formulate the ECVs' characteristics with a communication agent (CA) model, including the wireless communication coverage and mobility characteristics.

3) We formulate the coordination of ECVs, repair crews, and

switches with the interdependency constraints of CAs, repair agents (RAs), and electric agents (EAs).

It should be noted that, in our model, the usage of ECVs to set up emergency communications is to enable remote switching operations, rather than situational awareness. Also, the damage status is assumed to be known, which can be achieved by the damage assessment process, or by the outage information achieved via smart meters, customers' trouble calls, or even social sensors [20].

The rest of this paper is organized as follows. Section II introduces the integrated DSR framework, followed by the detailed formulations described in Section III. In Section IV, we give the solution methodology. Then, we test the proposed models by case studies and discuss the numerical results in Section V. In the end, we conclude this paper in Section VI.

## II. INTEGRATED DSR FRAMEWORK WITH CYBER-PHYSICAL INTERDEPENDENCE

In this section, we propose an integrated distribution system restoration framework as depicted in Fig. 1, which considers the emergency communication set up by ECVs where the existing communication infrastructure is ineffective. The control center sends commands for dispatching repair crews and ECVs and operating automatic switches remotely. The repair crews can repair damaged components and operate both automatic and manual switches. The bidirectional communication links between automatic feeder switches and the control center are built through the emergency wireless communication network set up by ECVs and FTUs. The seamless coordination of three sectors, i.e., the repair crews, the distribution system (physical sectors), and the emergency communication network (cyber sectors) are considered to help restore the unserved customers as quickly as possible. In the next section, we will introduce the individual and interdependent optimization models of these three sectors.

In this paper, we assume that in the short period (i.e. the scheduled time horizon), the existing communication infrastructure is not available and that the emergency wireless communication network can be set up by dispatching ECVs to certain working sites.

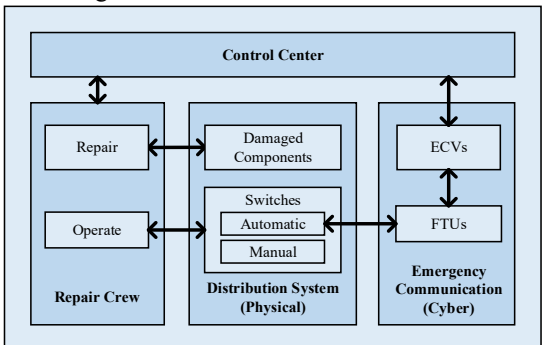

Fig. 1. Integrated distribution system restoration framework

## III. PROBLEM FORMULATIONS

In this section, we formulate and explain the proposed optimization models step by step.

### A. Emergency Communication Vehicle Model

An emergency communication vehicle (ECV) is actually a mobile base station. When it is dispatched to and set up at a certain working site (WS), an ECV and the available communication nodes (CNs) within its coverage can form a temporary wireless network. The working sites are certain locations given by expertized operators before dispatching the ECVs, in which many factors should be considered such as environment suitability for the erection of mobile base station, traffic accessibility for the vehicles, etc.

For a better explanation of the proposed ECV model, we use Fig. 2 to display the working process of ECVs. An ECV (e.g., $v_1, v_2$) departs from the depot where it is prepositioned. Then, it travels to a WS (e.g., $v_1$ to $k_1$, $v_2$ to $k_2$), and sets up a temporary base station. The ECV itself and the available CNs inside its cover range can quickly form a wireless network. As a result, the ECVs can transfer bidirectional signals between the control center and the communication nodes which are coupled with physical devices. In DSR, the CNs are the FTUs that are associated with automatic feeder switches. With the emergency wireless network, the communications between the control center and the FTUs can be built, realizing the resumed functionality of DA for DSR.

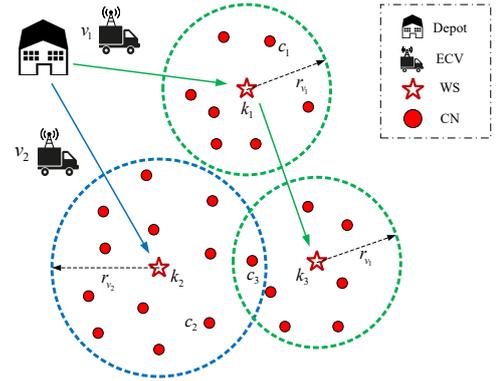

Fig. 2. Work process of emergency communication vehicles

An ECV can cover several CNs, and how many CNs can be covered is decided by its cover range. The cover range of an ECV $v$ is modeled by a circle denoted by its radius $r_v$. We use a binary parameter $u_{v,k,c}$ to represent if a communication node $c$ can be coved by the wireless network set up by ECV $v$ at the working site $k$, and it can be calculated given the original data, as expressed below:

$$u_{v,k,c} = \varepsilon\left(r_v - \sqrt{(x_c - x_k)^2 + (y_c - y_k)^2}\right),$$

where the step function: $\varepsilon(t) = \begin{cases} 1, t \geq 0 \\ 0, t < 0 \end{cases}$. For example, the CN $c_1$ is in the cover range of the base station set up by ECV $v_1$ at WS $k_1$, so the parameter $u_{v_1,k_1,c_1} = 1$. Similarly, the parameter $u_{v_1,k_1,c_2} = 0$, $u_{v_2,k_2,c_2} = 1$. It should be noted that a CN can be covered by different ECVs at different WSs, which are determined by their relative locations. For example, the CN $c_3$ can be both covered by ECV $v_1$ at WS $k_3$ and by ECV $v_2$ at WS $k_2$). Since the locations of CNs and WSs and the cover range of ECVs (i.e. $r_v$) are known, the $u_{v,k,c}$ are actually given parameters.

An ECV can travel from one WS to another, setting up a wireless network at another circle area to make some other intelligent physical devices observable and controllable. For example, in Fig. 2 ECV $v_1$ travels from WS $k_1$ to $k_2$. The mobility of an ECV is essentially a "vehicle routing" problem, which has been widely studied in operational research areas. Differently, for ECVs, the duration of stay at WSs is a variable, not a parameter, as the duration of stay depends on how long the emergency wireless network will be utilized, which is unknown in advance. To make it compatible with the models of repair crews and physical operational constraints of distribution systems, we adopt the "variable time step" (VTS) modeling method [11] to formulate the mobility of ECVs, and we add an extra variable $t_i^{Cd}$ (i.e., the departure time from WS $i$) so that the variable duration of stay can be modeled. Similar to [11], we use the concept of "communication agents" (CAs) to represent ECVs, and the specific constraints are formulated as below.

$$x_{ii}^C = 1, \forall i \in \mathcal{D}^C. \quad (1)$$
$$x_{ii}^C = 0, \forall i \in \mathcal{W}^C. \quad (2)$$
$$x_{ij}^C = 0, \forall i \in \mathcal{D}^C \cup \mathcal{W}^C, j \in \mathcal{D}^C, i \neq j. \quad (3)$$
$$x_{ij}^C + x_{ji}^C \leq 1, \forall i,j \in \mathcal{D}^C \cup \mathcal{W}^C. \quad (4)$$
$$\sum_{j=1,j\neq i}^{n^{CA}} x_{ij}^C \leq n_{cap,i}^C, \forall i \in \mathcal{D}^C. \quad (5)$$
$$\sum_{j=1}^{n^{CA}} x_{ij}^C \leq \sum_{h=1}^{n^{CA}} x_{hi}^C \leq 1, \forall i \in \mathcal{W}^C. \quad (6)$$
$$t_i^{Ca} = T_0, \forall i \in \mathcal{D}^C. \quad (7)$$
$$t_i^{Ca} \leq t_i^{Cd} \leq T^{MAX}, \forall i \in \mathcal{D}^C \cup \mathcal{W}^C. \quad (8)$$
$$\left.\begin{array}{l} t_j^{Ca} \geq t_i^{Cd} + T_{ij}^C - (1 - x_{ij}^C)M \\ t_j^{Ca} \leq t_i^{Cd} + T_{ij}^C + (1 - x_{ij}^C)M \end{array}\right\}, \forall i \in \mathcal{D}^C \cup \mathcal{W}^C, j \in \mathcal{W}^C. \quad (9)$$
$$t_j^{Cd} \geq t_j^{Ca} + T_j^{Cmin} - \left(1 - \sum_{i=1,i\neq j}^{n^{CA}} x_{ij}^C\right)M, \forall j \in \mathcal{W}^C. \quad (10)$$
$$\left.\begin{array}{l} t_j^{Ca} \geq T^{MAX} - M\sum_{i=1,i\neq j}^{n^{CA}} x_{ij}^C \\ t_j^{Ca} \leq T^{MAX} + M\sum_{i=1,i\neq j}^{n^{CA}} x_{ij}^C \end{array}\right\}, \forall j \in \mathcal{W}^C. \quad (11)$$

Constraints (1–6) define the route table of communication agents (CAs). Specifically, (1–3) mean that a CA should travel starting only from depots and should not go back to depots. Constraint (4) represents that each possible route can be visited no more than once because the routing of ECVs is coordinated with the service restoration process so that at the visited working sites, there should always be a "task" of setting up emergency communications for remotely switching, and after the task is completed, there should be no further need for communication, in other words, there is no need for ECVs to revisit this working site. Constraint (5) limits the total number of agents dispatched out of the depot not to exceed the capacity of that depot. Besides, (6) depicts that each WS can be visited by at most one CA, and one CA leave or stay at the visited WS. Except for route table-related constraints, (7–11) list the time-related constraints. Specifically, (7) defines the initial time of a CA in the time horizon. Constraint (8) ensures that, at any site (depot or WS), the departure time should be later than the arrival time, and both times should not exceed the scheduled time horizon. Constraint (9) is tight only if there is a CA travels from $i$ to $j$ (i.e., the route table element $x_{ij}^C = 1$), which limits the time difference between arriving at site $j$ and departing from site $i$ to be exactly the travel time from $i$ to $j$. Constraint (10) ensures the duration of stay at a WS $j$ should be longer than the required minimum duration if one CA visits this site. Constraint (11) sets the arrival time at a WS to be the end time of the scheduled horizon (i.e. $T^{MAX}$) if there are no CAs traveling to that WS.

*B. Crew Dispatch Model*

To reduce the complexity of the crew dispatch model, first, we pre-assign working sites (WSs) of repair crews (including switches and faulted lines to be repaired) to depots. Similar to [21], the clustering model is formulated as follows:

$$min \quad \sum_{i\in\mathcal{D}^R}\sum_{j\in\mathcal{W}^R} y_{ij} T_{ij}^R \quad (p1)$$
$$s.t. \quad \sum_{i\in\mathcal{D}^R} y_{ij} = 1, \forall j \in \mathcal{W}^R \quad (p2)$$

By solving the above integer programming problem, the optimal clustering strategy with minimum travel time can be found. Then, we use the "repair agent" (RA) to represent the repair crew, which can be modeled by the constraints of the route table and arrival time list [11], which are listed below.

$$x_{ii}^R = 1, \forall i \in \mathcal{D}^R \quad (12)$$
$$x_{ii}^R = 0, \forall i \in \mathcal{W}^R \quad (13)$$
$$x_{ij}^R = 0, \forall i,j \in \mathcal{D}^R, i \neq j \quad (14)$$
$$x_{ij}^R + x_{ji}^R \leq 1, \forall i,j \in \mathcal{D}^R \cup \mathcal{W}^R \quad (15)$$
$$\sum_{j=1,j\neq i}^{n^R} x_{ij}^R \leq n_{cap,i}^R, \forall i \in \mathcal{D}^R \quad (16)$$
$$\sum_{j=1}^{n^R} x_{ij}^R \leq \sum_{h=1}^{n^R} x_{hi}^R \leq 1, \forall i \in \mathcal{W}^R \quad (17)$$
$$t_i^R = T_0, \forall i \in \mathcal{D}^R \quad (18)$$
$$\left.\begin{array}{l} t_j^R \geq t_i^R + T_{ij}^R - (1-x_{ij}^R)M \\ t_j^R \leq t_i^R + T_{ij}^R + (1-x_{ij}^R)M \end{array}\right\}, \forall i \in \mathcal{D}^R, \forall j \in \mathcal{W}^R. \quad (19)$$
$$\left.\begin{array}{l} t_j^R \geq t_i^R + T_i^{RP} + T_{ij}^R - (1-x_{ij}^R)M \\ t_j^R \leq t_i^R + T_i^{RP} + T_{ij}^R + (1-x_{ij}^R)M \end{array}\right\},$$
$$\forall i \in \mathcal{F}\backslash\mathcal{SW}, \forall j \in \mathcal{W}^R, i \neq j. \quad (20)$$
$$\left.\begin{array}{l} t_j^R \geq t_i^R + T_i^{MS} + T_{ij}^R - (1-x_{ij}^R)M \\ t_j^R \leq t_i^R + T_i^{MS} + T_{ij}^R + (1-x_{ij}^R)M \end{array}\right\},$$
$$\forall i \in \mathcal{SW}\backslash\mathcal{F}, \forall j \in \mathcal{W}^R, i \neq j. \quad (21)$$
$$\left.\begin{array}{l} t_j^R \geq t_i^R + T_i^{RP} + T_i^{MS} + T_{ij}^R - (1-x_{ij}^R)M \\ t_j^R \leq t_i^R + T_i^{RP} + T_i^{MS} + T_{ij}^R + (1-x_{ij}^R)M \end{array}\right\},$$
$$\forall i \in \mathcal{SW} \cap \mathcal{F}, \forall j \in \mathcal{W}^R, i \neq j. \quad (22)$$
$$\left.\begin{array}{l} t_j^R \geq T^{MAX} - M\sum_{i=1,i\neq j}^{n^{RA}} x_{ij}^R \\ t_j^R \leq T^{MAX} + M\sum_{i=1,i\neq j}^{n^{RA}} x_{ij}^R \end{array}\right\}, \forall j \in \mathcal{W}^R. \quad (23)$$
$$f_{r^e(f)}^R \geq t_f^R + T_f^{RP}, \forall f \in \mathcal{F}. \quad (24)$$
$$f_i^R = T_0, \forall i \in \mathcal{C}^E, i \neq r^e(f), f \in \mathcal{F}. \quad (25)$$

In this paper, we assume all the crews have both skills of operating switches and repairing faulted lines, thus the operation agent (OA) and repair agent (RA) in [11] are unified to be only RA. In other words, the working sites of RAs include both switches and faulted lines, i.e. $\mathcal{W}^R = \mathcal{SW} \cup \mathcal{F}$. First, the RAs' route table-related constraints are listed in (12)–(17). Specifically, (12)–(13) limit that a RA should travel starting only from the depot. Constraint (14) limits that each type of agent should not go back to the depot because how a RA travel back to the depot from the last visited site is irrelevant to the proposed problem. Constraint (15) means each possible route can be visited no more than once because each working site cannot be visited twice. Constraint (16) means the total number of RAs dispatched out of each depot cannot exceed the capacity of that depot. Constraint (17) describes that each working site can be visited by at most one RA, and a RA should leave or stay at the visited working site. Second, the time-related constraints of RAs are listed in (18–25). Specifically, (18) defines the initial time of RAs in the scheduled time horizon. Constraints (19–22) describe the equality of arrival time between two sites if a RA travels from one site to another. These constraints include: from a depot to a WS (i.e. (19)), from a faulted line (not switch) to another WS (i.e. (20)), from a healthy switch to another WS (i.e. (21)), and from a faulted switch to another WS (i.e. (22)). Constraint (23) sets the arrival time of a WS to be the end time of the scheduled horizon (i.e. $T^{MAX}$) if there are no RAs travel to that WS. Constraints (24–25) limit the repair completion time of different types of node cells, where $r^e(f)$ is the index transfer from RA to EA, which represents the node cell where the fault $f$ is inside. For a node cell with faulted lines inside it, the repair completion time of this cell is later than that of all the faulted lines inside it, as shown in (24). By contrast, for the node cell without faulted lines inside it, the repair completion time is set to be the initial time of RAs in the scheduled horizon, as shown in (25).

*C. Physical System Model*

We use the variable time step (VTS) modeling method, as introduced in [4], to formulate the physical distribution systems. The virtual electric agents (EA) are used to represent the energy flow in the physical network, which departs from a substation or black-start DG and goes through node cells to restore unserved load in these cells. Specifically, the physical system models include 1) the constraints of EAs' route table; 2) the constraints of EAs' arrival time; 3) the constraints of energization status; 4) the constraints of system and component operation. All these constraints can be found in [4], and we integrate them labeled by one number, as summarized below.

TABLE I. THE INTEGRATED PHYSICAL SYSTEM MODEL

| Number in this paper | Number in [4] | Description |
|---|---|---|
| (E26) | (1)–(6) | Constraints of the route table of EAs |
|  | (7)–(9) | Constraints of the arrival time of EAs |
|  | (12)–(22) | Constraints of the nodes', cells', and components' energization status and their relationship |
|  | (23)–(31), (37)–(42) | Constraints of the system operation (three-phase voltage and power equations), and the components' operation (DGs, regulators, lines, and loads) |

*D. Interdependency Constraints*

In the abovementioned models in section III.A–C, we use CAs, RAs, and EAs to represent the emergency communication vehicles, the repair crews, and the physical distribution systems, respectively. In this section, we list all the interdependent constraints among these agents.

*1) CA-EA (Cyber-Physical) Interdependency Constraints*

For healthy automatic switches, they can either be closed remotely by DA or manually by repair crews, which can be expressed as:
$$x_{ij}^E = d_{ij}^{AO} + d_{ij}^{MO}, \forall (i,j) \in \mathcal{AS}\backslash\mathcal{F}. \quad (27)$$

Basically, two conditions should be satisfied to enable the feeder automation to remotely close the opened switches: 1) the communication between the control center and the FTUs is unblocked to enable information flows; 2) the FTUs have power sources for the automatic control. According to the introduction in section III.A, when an ECV is dispatched to and set up at a WS, all the operable FTUs inside the cover range of the wireless network set up by the ECV can restore communications to the control center, making the associated automatic switches remotely operable again. Note that the FTUs are equipped with backup batteries with limited capacity, so the "residual time" of an FTU, denoted as $RT_{(i,j)'}$ indicating the duration of depleting the backup battery of an FTU, should also be considered. We formulate the cyber-physical interdependency between CA and EA as below.

$$d_{ij}^{AO} + d_{ji}^{AO} = \sum_{k \in \mathcal{W}_{(i,j)'}^C} z_{(i,j)',k}^C \quad (28)$$

$$\sum_{k \in \mathcal{W}_{(i,j)'}^C} z_{(i,j)',k}^C \leq 1 \quad (29)$$

$$z_{(i,j)',k}^C = 0, k \notin \mathcal{W}_{(i,j)'}^C \quad (30)$$

$$z_{(i,j)',k}^C \leq \sum_{h=1, h \neq k}^{n^{CA}} x_{hk}^C, k \in \mathcal{W}_{(i,j)'}^C \quad (31)$$

$$0 \leq t_{ij}^{AOop} \leq T^{MAX} \quad (32)$$

$$\left.\begin{array}{l} t_{ij}^{AOop} \geq t_k^{Ca} - (2 - z_{(i,j)',k}^C - d_{ij}^{AO})M \\ t_{ij}^{AOop} \leq t_k^{Cd} + (2 - z_{(i,j)',k}^C - d_{ij}^{AO})M \end{array}\right\} \quad (33)$$

$$\left.\begin{array}{l} d_{ij}^{AOe} + d_{ij}^{AOde} \geq 1 - (1 - d_{ij}^{AO})M \\ d_{ij}^{AOe} + d_{ij}^{AOde} \leq 1 + (1 - d_{ij}^{AO})M \end{array}\right\} \quad (34)$$

$$\frac{t_{ij}^{AOop} - \max(t_i^E, f_j^R)}{M} \leq d_{ij}^{AOe} \leq \frac{t_{ij}^{AOop} - \max(t_i^E, f_j^R)}{M} + 1 \quad (35)$$

$$\left.\begin{array}{l} t_j^E \geq t_{ij}^{AOop} + T_{(i,j)}^{AS} - (2 - d_{ij}^{AO} - d_{ij}^{AOe})M \\ t_j^E \leq t_{ij}^{AOop} + T_{(i,j)}^{AS} + (2 - d_{ij}^{AO} - d_{ij}^{AOe})M \end{array}\right\} \quad (36)$$

$$RT_{(i,j)'} \geq t_{ij}^{AOop} - (2 - d_{ij}^{AO} - d_{ij}^{AOe})M, (i,j)' \text{at } j \quad (37)$$

$$\frac{t_i^E - (t_{ij}^{AOop} + T_{(i,j)}^{AS})}{M} \leq d_{ij}^{AOde} \leq \frac{t_i^E - (t_{ij}^{AOop} + T_{(i,j)}^{AS})}{M} + 1 \quad (38)$$

$$\left.\begin{array}{l} t_j^E \geq t_i^E - (2 - d_{ij}^{AO} - d_{ij}^{AOde})M \\ t_j^E \leq t_i^E + (2 - d_{ij}^{AO} - d_{ij}^{AOde})M \end{array}\right\} \quad (39)$$

$$RT_{(i,j)'} \geq t_{ij}^{AOop} - (2 - d_{ij}^{AO} - d_{ij}^{AOde})M \quad (40)$$

Constraints (28–40) give the conditions that a healthy automatic switch $(i,j) \in \mathcal{AS}\backslash\mathcal{F}$ can be operated remotely, where $(i,j)'$ is the FTU at the switch $(i,j)$, embedded with communication and control devices. Constraint (28–29) means that if and only if there is a mobile base station at the working site that can cover the CN $(i,j)'$, then the switch $(i,j)$ can be operated automatically (either from $i$ to $j$ or from $j$ to $i$), and $(i,j)'$ can only be governed by at most one working site. Constraints (30–31) limit the variable $z_{(i,j)',k}^C$ by considering the cover range and CAs' routing behaviors. Specifically, (30) indicates that the base station at WS $k$ cannot supply the communication for the automatic switch $(i,j)$ beyond the cover range of the ECV at WS $k$. Constraint (31) indicates that the FTU $(i,j)'$ cannot be governed by WS $k$ if there are no ECVs visit this WS. Constraints (32–40) descript the relationship between the remote operation time of an automatic switch ($t_{ij}^{AOop}$), the arrival time and depart time of CAs at a WS ($t_k^{Ca}, t_k^{Cd}$), and the energization time of node cells on both sides

$(t_i^E, t_j^E)$. Specifically, (32) limits the range of the remote operation time. Constraint (33) indicates that the remote operation time of the automatic switch $(i,j)$ should be within the time interval $[t_k^{Ca}, t_k^{Cd}]$ if the FTU $(i,j)'$ is governed by the ECV at WS $k$ (i.e. $z_{(i,j)',k}^C = 1$) and this switch is automatically operated (i.e. $d_{ij}^{AO} = 1$). Constraint (34) indicates that if the switch $(i,j)$ is automatically operated, then it should be either energized or de-energized operated, followed by the constraints (35–37) which give the conditions of energized operation, and the constraints (38–40) which give the conditions of de-energized operation, as introduced in the following two paragraphs, respectively.

In (35), $\max(t_i^E, f_j^R)$ represent the earliest time when the automatic switch $(i,j)$ is "ready" to be switched on from node cell $i$ to $j$, and we define it as the "ready time". It consists of two conditions that must be satisfied if the automatic switch $(i,j)$ is ready to be switched on from node cell $i$ to $j$: 1) the node cell $i$ has been energized (i.e. after $t_i^E$), and 2) all the faults in the node cell $j$ have been repaired (i.e. after $f_j^R$). For energized operation from $i$ to $j$ (i.e. $d_{ij}^{AOe} = 1$), the remote operation time from $i$ to $j$ for automatic switch $(i,j)$ (i.e. $t_{ij}^{AOop}$) should be later than the ready time "$\max(t_i^E, f_j^R)$", as indicated in (35). In this situation, the node cell $j$ will be energized immediately after the automatic switch $(i,j)$ is remotely operated or closed, as depicted in (36). After outages caused by extreme events, the power supply of an FTU from the power grid is lost, and the backup battery of the FTU continues to supply it. It should be noted that the power supply of FTU $(i,j)'$ from the power grid is only at one side of the switch $(i,j)$. If the FTU $(i,j)'$ is at the "from" node cell $i$ and the automatic switch $(i,j)$ is remotely operated from $i$ to $j$, then the power source of FTU is not a concern because the grid can supply the power. However, if the FTU $(i,j)'$ is at $j$ side, then the remote operation time of the automatic switch $(i,j)$ must be before the residual time of $(i,j)'$, as described in (37).

In (38), "$t_{ij}^{AOop} + T_{(i,j)}^{AS}$" represent the remote operation completion time of the automatic switch $(i,j)$. For de-energized operation (i.e. $d_{ij}^{AOde} = 1$), the energization time of both node cells at two ends of the automatic switch $(i,j)$ should be the same and later than the operation completion time, as depicted in (38–39). Meanwhile, since both sides are de-energized during the operation, the backup battery is the only power source of FTU and the switch's remote operation. Thus, in this situation, the remote operation time of the automatic switch $(i,j)$ must be before the residual time of $(i,j)'$ no matter which side of FTU $(i,j)'$ is at, as depicted in (40).

2) *RA-EA Interdependent Constraints*

Any switch ($\forall (i,j) \in \mathcal{SW}$) can be closed manually, and the following constraints list the interdependent constraints between RA and EA when a switch is manually operated. Note that, in this paper, we assume all the repair crews can operate energized or de-energized switches.

$$d_{ij}^{MO} + d_{ji}^{MO} = \sum_{h=1, h\neq k}^{n^{RA}} x_{hk}^R, \forall (i,j) \in \mathcal{SW}, k = e^r(i,j). \quad (41)$$

$$\left.\begin{array}{l} d_{ij}^{MOe} + d_{ij}^{MOde} \geq 1 - (1 - d_{ij}^{MO})M \\ d_{ij}^{MOe} + d_{ij}^{MOde} \leq 1 + (1 - d_{ij}^{MO})M \end{array}\right\}, \forall (i,j) \in \mathcal{SW}. \quad (42)$$

$$\frac{t_{e^r(i,j)}^R - t_i^E}{M} \leq d_{ij}^{MOe} \leq \frac{t_{e^r(i,j)}^R - t_i^E}{M} + 1, \forall (i,j) \in \mathcal{SW}. \quad (43)$$

$$\left.\begin{array}{l} t_j^E \geq t_{e^r(i,j)}^R + T_{(i,j)}^{MS} - (2 - d_{ij}^{MO} - d_{ij}^{MOe})M \\ t_j^E \leq t_{e^r(i,j)}^R + T_{(i,j)}^{MS} + (2 - d_{ij}^{MO} - d_{ij}^{MOe})M \end{array}\right\},$$
$$\forall (i,j) \in \mathcal{SW} \backslash \mathcal{F}. \quad (44)$$

$$\left.\begin{array}{l} t_j^E \geq t_{e^r(i,j)}^R + T_{(i,j)}^{RP} + T_{(i,j)}^{MS} - (2 - d_{ij}^{MO} - d_{ij}^{MOe})M \\ t_j^E \leq t_{e^r(i,j)}^R + T_{(i,j)}^{RP} + T_{(i,j)}^{MS} + (2 - d_{ij}^{MO} - d_{ij}^{MOe})M \end{array}\right\},$$
$$\forall (i,j) \in \mathcal{SW} \cap \mathcal{F}. \quad (45)$$

$$\frac{t_i^E - (t_{e^r(i,j)}^R + T_{(i,j)}^{MS})}{M} \leq d_{ij}^{MOde} \leq \frac{t_i^E - (t_{e^r(i,j)}^R + T_{(i,j)}^{MS})}{M} + 1,$$
$$\forall (i,j) \in \mathcal{SW} \backslash \mathcal{F}. \quad (46)$$

$$\frac{t_i^E - (t_{e^r(i,j)}^R + T_{(i,j)}^{RP} + T_{(i,j)}^{MS})}{M} \leq d_{ij}^{MOde} \leq \frac{t_i^E - (t_{e^r(i,j)}^R + T_{(i,j)}^{RP} + T_{(i,j)}^{MS})}{M} + 1,$$
$$\forall (i,j) \in \mathcal{SW} \cap \mathcal{F}. \quad (47)$$

$$\left.\begin{array}{l} t_j^E \geq t_i^E - (2 - d_{ij}^{MO} - d_{ij}^{MOde})M \\ t_j^E \leq t_i^E + (2 - d_{ij}^{MO} - d_{ij}^{MOde})M \end{array}\right\}, \forall (i,j) \in \mathcal{SW}. \quad (48)$$

In these constraints, $e^r(i,j)$ denotes the index transfer from EA to RA, which represents the working site for repair crews to repair and operate the switch $(i,j)$. Constraint (41–42) implies that if the switch $(i,j)$ is visited by a RA, then it must be manually closed either from $i$ to $j$ or from $j$ to $i$, and be either energized or de-energized during the operation. The constraints of energized and de-energized operation of the switch $(i,j)$ are expressed in (43–45) and (46–48), respectively. For the energized operation of the switch $(i,j)$ from $i$ to $j$ to energize the node cell $j$, the node cell $i$ has been restored before a RA arrives at it, as shown in (43). In this case, the other node cell $j$ will be restored immediately after the RA has switched on $(i,j)$ if $(i,j)$ is healthy, or after the RA has repaired $(i,j)$ and switched on it if it needs to be repaired, as shown in (44–45). For the de-energized operation of the switch $(i,j)$ from $i$ to $j$, the node cell $i$ can only be energized after the switch $(i,j)$ has been repaired and closed, as shown in (46–47). In this case, both the node cell $i$ and $j$ will be restored immediately at the same time, as depicted in (48).

Except for manual operation constraints, we have additional RA-EA interdependent constraints listed below.

$$\left.\begin{array}{l} t_i^E \geq t_{e^r(i,j)}^R + T_{(i,j)}^{RP} \\ t_j^E \geq t_{e^r(i,j)}^R + T_{(i,j)}^{RP} \end{array}\right\}, \forall (i,j) \in \mathcal{SW} \cap \mathcal{F}. \quad (49)$$

$$t_i^E \geq f_i^R, \forall i \in \mathcal{C}^E. \quad (50)$$

Constraint (49) ensures both end cells of a faulted switch can only be energized after it is repaired. Constraint (50) limits the restoration time of any node cell must be after the repair completion time.

*E. Objective Functions*

We define the objective functions as below.

$$Obj^{EA} = \frac{\sum_{c \in \mathcal{C}^L} \omega_c^{EA} t_c^{EA}}{\sum_{c \in \mathcal{C}^L} \omega_c^{EA} T^{MAX}} \quad (51)$$

$$Obj^{RA} = \frac{\omega_1^{RA} \sum_{i=1}^{n^{RA}} \sum_{j=1, j\neq i}^{n^{RA}} x_{ij}^R T_{ij}^R}{n^{RA} T^{MAX}} + \frac{\omega_2^{RA} \sum_{j=1, j \notin \mathcal{DR}}^{n^{RA}} \left(\sum_{i=1, i\neq j}^{n^{RA}} x_{ij}^R\right)(t_j^R + T_j^{MS})}{n^{RA} T^{MAX}} \quad (52)$$

$$Obj^{CA} = \frac{\omega_1^{CA}\sum_{i=1}^{n^{CA}}\sum_{j=1,j\neq i}^{n^{CA}} x_{ij}^C T_{ij}^C}{n^{CA} T^{MAX}} + \frac{\omega_2^{CA}\sum_{i\in\mathcal{W}^C}(t_i^{Cd}-t_i^{Ca})}{n^{CA} T^{MAX}} \quad (53)$$

The proposed multi-objective functions include the EA, RA, and CA-related normalized objectives, as shown in (51–53) respectively. The $Obj^{EA}$ in (51) is the EA-related objective, which represents the ratio (in percentage) of the weighted loads' unserved time to the scheduled time horizon ($\omega_c^{EA}$ represent the weight of load cell $c$). The $Obj^{RA}$ in (52) is the RA-related objective, which includes two parts: the first part is the total travel time of all the RAs, and the second part is the total working time (including the repair time for damaged components and the manual operation time for switches) of all the RAs; both are normalized with the number of RAs multiplied by the scheduled time horizon, and they are multiplied by different weights (i.e. $\omega_1^{RA}$ and $\omega_2^{RA}$, respectively). The $Obj^{CA}$ in (53) is the CA-related objective, which also includes two parts: the first part is the total travel time of all the CAs, and the second part is the total working time at sites (i.e. the duration of stay at working sites); both are normalized with the number of CAs multiplied by the scheduled time horizon, and they are multiplied by different weights (i.e. $\omega_1^{CA}$ and $\omega_2^{CA}$, respectively).

It can be found that: 1) the value of $Obj^{EA}$ represents the average unserved time of each load cell, which should be minimized; 2) the value of $Obj^{RA}$ or $Obj^{CA}$ represent the average time (including travel time and working time) that a RA or CA takes to coordinate with the restoration, which should also be minimized. In (51–53), all the coefficients ($\omega$ with superscript) are weights given by the decision-makers.

*F. The Whole Optimization Models*

The integrated distribution system restoration optimization models are categorized into two types, according to whether the service restoration is with CAs (i.e. ECVs) or without CAs, which are respectively denoted as "OPT-WCA" and "OPT-WOCA", i.e.:

**OPT-WCA**:
$$Min. Obj^{sum} = \beta^{EA} Obj^{EA} + \beta^{RA} Obj^{RA} + \beta^{CA} Obj^{CA} \quad (54)$$
$s.t.$ CA's constraints: (1–11)
RA's constraints: (12–25)
EA's constraints: (E26)
Interdependent constraints: (27–50)
Objectives: (51–53)

**OPT-WOCA:**

First, constraint (27) should be revised to adapt to the situation that all the switches can only be operated manually, as shown below:
$$x_{ij}^E = d_{ij}^{MO}, \forall (i,j) \in \mathcal{SW}. \quad (R27)$$

Then, the whole optimization model can be listed below:
$$Min. Obj^{sum} = \beta^{EA} Obj^{EA} + \beta^{RA} Obj^{RA} \quad (55)$$
$s.t.$ RA's constraints: (12–25)
EA's constraints: (E26)
Interdependent constraints: (R27), (41-50)
Objectives: (51–52)

As introduced in Section III.E, in order to make the values of the three objectives (i.e. $Obj^{EA}, Obj^{RA}$ and $Obj^{CA}$) comparable, they are normalized and have the same units (in p.u.). In (54) and (55), the coefficients ($\beta$ with superscript) are the weights of these three objectives, which are chosen by the decision-makers according to their importance. In this paper we set $\beta^{EA}:\beta^{RA}:\beta^{CA} = 10:1:1$, because we think restoring unserved loads from outages is much more important and urgent than the cost-saving of emergency resources (including repair crews and ECVs). By solving these two optimization problems and comparing the optimization results, we can see the benefits of ECVs for distribution system restoration.

*G. Additional Discussion*

After extreme events, the communication network may be fully unavailable, or some of them may be intact. In the former scenario, the feeder automation and remote operation of automatic switches can only rely on emergency communication set up by temporary base stations such as ECVs proposed in this paper. The proposed model above can handle such a situation. In the latter scenarios, the communication network may be partially damaged. The proposed model can also handle such a situation by fixing some conditions, as introduced below.

First, if a base station at a working site $k$ is intact, we exclude $k$ from the set of ECVs' work sites, i.e. $\mathcal{W}^{C'} = \mathcal{W}^C\setminus\{k\}$. Second, for an automatic switch $(i,j)$ and the corresponding FTU $(i,j)'$ within the cover range of the base station at $k$:

1) If both the automatic switch $(i,j)$ and the FTU $(i,j)'$ are intact, then $(i,j)$ can be operated remotely without ECVs at any time in the scheduled horizon, and there is no need to dispatch a repair crew to operate it. We can handle this situation by the following steps: step 1, we exclude switch $(i,j)) \in \mathcal{AS}\setminus\mathcal{F}$ from the working site of repair crews, i.e. $\mathcal{W}^{R'} = \mathcal{W}^R\setminus\{(i,j)\}$; step 2, we fix constraints (28) and (41) to be equalities, i.e. $d_{ij}^{AO}+d_{ji}^{AO}=1$ and $d_{ij}^{MO}+d_{ji}^{MO}=0$; step 3, we exclude constraints (29–31) and (33).

2) If either the automatic switch $(i,j)$ or the FTU $(i,j)'$ is damaged, then $(i,j)$ cannot be controlled remotely and can only be manually operated by a repair crew. We can handle this situation by considering $(i,j) \in \mathcal{AS}\cap\mathcal{F}$, of which the repair time $T_{(i,j)}^{RP}$ is the real repair time if the switch $(i,j)$ is damaged and zero if FTU $(i,j)'$ is damaged because repairing FTUs is not that urgent compared with operating switches during the DSR after large-scale outages.

IV. SOLUTION METHODOLOGY

The proposed optimization models (i.e. "OPT-WCA" and "OPT-WOCA") are mixed-integer programming (MIP) problems, of which all the objective functions and constraints are linear except for the nonlinear terms max(·) in (35). By linearizing these nonlinear terms, the whole models are transferred into mixed-integer linear programming (MILP) problems which can be effectively solved by off-the-shelf solvers such as Cplex and Gurobi.

The maximum value of two variables ($y = \max\{x_1, x_2\}$) can be linearized by introducing two binary variables ($d_1, d_2$) [22], and the equivalent MILP formulations are listed below:
$$L_i \leq x_i \leq U_i, \forall i = 1,2$$
$$y \geq x_i, \forall i = 1,2$$
$$y \leq x_i + (U_{max} - L_i)(1-d_i), \forall i = 1,2$$
$$d_1 + d_2 = 1$$

where $L_i$ and $U_i$ are lower and upper bound of the variable $x_i$, and $U_{max}$ is the maximum value of all the upper bounds. In our problems, according to the definition the variables $t_i^E$ and $f_j^R$ must be within the scheduled time horizon ($[0, T^{MAX}]$). Thus, we use an auxiliary variable $y_{ij}$ to replace the nonlinear term (i.e. $\max\{t_i^E, f_j^R\}$) in (35) and introduce two binary variables $d_{ij}^E$ and $d_{ij}^R$ to formulate the equivalent MILP constraints, as listed below:

$$y_{ij} \geq t_i^E \quad (A.1)$$
$$y_{ij} \geq f_j^R \quad (A.2)$$
$$y_{ij} \leq t_i^E + T^{MAX}(1 - d_{ij}^E) \quad (A.3)$$
$$y_{ij} \leq f_j^E + T^{MAX}(1 - d_{ij}^R) \quad (A.4)$$
$$d_{ij}^E + d_{ij}^R = 1 \quad (A.5)$$

where $\forall i,j \in \mathcal{C}^E, (i,j) \in \mathcal{AS}\backslash\mathcal{F}$. By replacing the nonlinear term $\max\{t_i^E, f_j^R\}$ with the auxiliary variable $y_{ij}$ and adding the auxiliary constraints (A.1–A.5) to the proposed "OPT-WCA" model, the problems become MILP models.

## V. CASE STUDY

In this section, we test the proposed optimization models on IEEE 123 node test feeder, solved by Gurobi 9.5.2 on a PC with Intel Core i7-7500U 2.90-GHz CPU, 16-GB RAM, and 64-bit operating system.

### A. Case Design and Parameters

We use the 123 node test feeder, which is a medium-size unbalanced distribution system operating at 4.16 kV nominal voltage with 3385 kW three-phase unbalanced loads in total [23]. The one-line diagram of the test system located in a rectangular coordinate system is shown in Fig. 3, in which all the nodes and lines are marked in grey and all the switches are open, which means they are all de-energized at the beginning of the scheduled horizon. Also, we assume the substation is initially unavailable and can start to supply power at the 30th minute.

As shown in Fig. 3, we have allocated 4 repair crews at 2 depots (2 crews at each one) to be prepared to visit 20 candidate working sites, including 4 faulted lines and 16 switches. By solving the integer programming problem (p1–p2), the 4 faulted lines are clustered to 2 depots, in which (13, 34) and (47, 48) belong to depot D1, (76, 77) and (101, 102) belong to depot D2. For the switches, we assume all of them are automatic switches installed with FTUs (labeled with red solid dots on the top of the switches) which can be communicated with and controlled remotely. Besides, we assume all the switches could be either closed remotely through feeder automation or closed manually by repair crews. As for the cyber part, we assume the existing communication network is unavailable, which means all the automatic switches are not able to be operated remotely if there is no emergency communication.

In this section, we design four cases to validate the proposed models. In Case 1 and Case 2, we assume there are no ECVs, and the repair crews can both repair faulted lines and operate switches. For Case 1, we use a simple heuristic rule to dispatch repair crews, labeled as *Algorithm 1* and depicted in Table II. It can be found that *Algorithm 1* is in favor of finding the optimal $Obj^{RA}$ in (54) for a given electric route. For Case 2, we use the proposed "OPT-WOCA" model to optimize the route and sequence of repair crews. For the convenience of comparison, we use the electric path in Case 2's results to be the given electric path which is normally supplied by DSO.

In Case 3 and Case 4, we have 2 ECVs at the 2 depots (1 ECV at each depot). The ECVs are prepared to visit 6 candidate working sites to set up an emergency wireless communication network, in which the cover ranges of these 2 ECVs are the same with a radius of 10 units, as labeled with dashed circles in Fig. 3. For case 3, we dispatch ECVs to WSs that can cover the largest number of FTUs, and they do not go to other WSs. This heuristic rule can be named the "***Maximum Coverage Algorithm***", which is commonly used in real-world communication recovery practices. In our proposed model, this algorithm can be realized only by fixing the CA's route table. It should be noted that, in this case, the interdependencies among CAs, EAs, and RAs are also considered by solving the proposed "OPT-WCA" model, which differs from only considering communication recovery. For case 4, we use the proposed "OPT-WCA" model to co-optimize the sequence of CAs, RAs, and EAs, and the interdependencies among them.

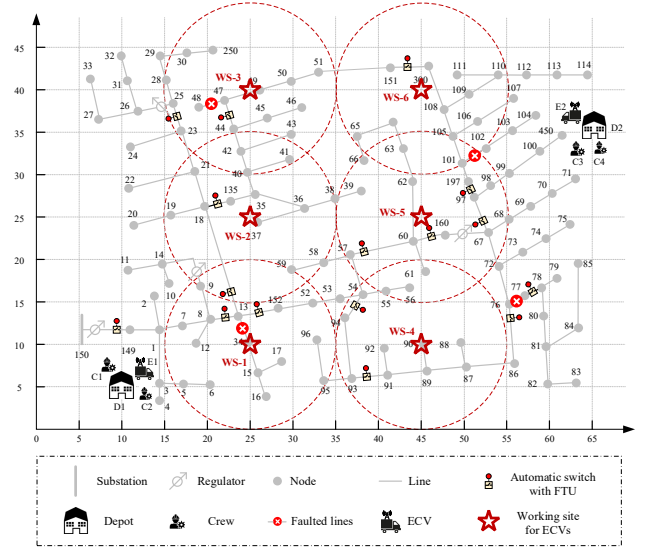

Fig. 3. The 123 node test feeder modified for case studies

TABLE II. A HEURISTIC CREW DISPATCH RULE

| *Algorithm 1*: Crew Dispatch by A Heuristic Rule |
|---|
| ***Step 1***: Form the set of working sites to be visited by repair crews ($\mathcal{W}^R$): according to the electric path given by the distribution system operator (DSO), decide the faulted lines and switches that are to be visited by repair crews. |
| ***Step 2***: Cluster work sites ($\mathcal{W}^R$) to depots ($\mathcal{D}^R$) by solving the integer optimization model (p1–p2). |
| ***Step 3***: Cluster work sites in cluster ($\mathcal{W}_d^R, d \in \mathcal{D}^R$) to crews at each depot by solving an integer optimization model similar to (p1–p2). |
| ***Step 4***: For each crew: first, visit and repair the faulted lines within its cluster; then, operate switches step by step. In each crew's cluster, the visiting sequence is choosing the site that is closest to the current site, until all the sites are visited. |
| ***Step 5***: Calculate the repair completion time of faulted lines, operation time of switches, and energization time of node cells. |

As for the weights of multi-objectives in (54–55), since the most important and urgent task is restoring unserved loads as quickly as possible, we set the weights of different agents to be: $\beta^{EA}=10, \beta^{CA}=\beta^{RA}=1$. Besides, we set all the $\omega$ weights in (51–53) to be 1.

For other parameters, we set: 1) the time horizon: 12 hours; 2) the repair time of faulted lines: 2 hours for each one; 3) the travel time for repair crews and ECVs traveling between two sites: be proportional to the Euclid distances in Fig. 3, in which the maximum is 65 minutes between depot D1 and depot D2; 4) switching time for remote operation: 1 minute; 5) switching time for manual operation: 15 minutes; and 6) the residual time for FTUs: 4 hours for each one; 7) the minimum duration of stay of ECVs at each WS: 15 minutes. All the optimization models are solved by Gurobi 9.5.2 because Gurobi has well-known advantages in solving MIP problems compared to Cplex. The solver is set to be with a relative MIP Gap tolerance of 0.001 and a "TimeLimit" of 900 seconds (i.e. 15 minutes).

*B. Results and Discussions*

The optimization results of four cases are listed in Table III. It can be found that: 1) the computation time of Case 1 is much less than that of Case 2 because *Algorithm 1* is a heuristic rule-based method and no complex optimization models are used; 2) the computation time of Case 3 is much less than Case 4 because the route table of ECVs is fixed which greatly reduces the computation complexities of the proposed "OPT-WCA" model. It can also be found that the MIP gaps of Cases 2 and 4 are larger than the given tolerance within the given computation time limit. However, by observing the Gurobi MIP solution logs of Case 2 and 4, we find that the incumbent objective values remain unchanged from the early stage of the computation (the 73$^{rd}$ second for Case 2 and the 177$^{th}$ second for Case 4) and that the best lower bounds (because our optimization models are minimization problems) approach the incumbent objective values slower and slower over time. Taking Case 4 as an example, this trend can be easily found in Fig. 4(a) – the MIP objective bounds and Fig. 4(b) – the MIP Gap during the Gurobi's solution process. Just as Ahmadi [24] analyzed, the program reaches the real optimal solution quickly, but it takes a long time to prove the optimality, so defining a limit for the solution time is an effective way to substantially reduce solution time. Based on the abovementioned analysis, for Cases 2 and 4, it is recommended to set the Gurobi's time limit to 180 seconds (i.e. 3 minutes) to save the computation time while ensuring the solutions' quality.

As shown in Table III, the EA, RA, and CA-related objective values are expressed in percentages, of which the meanings are explained in Section III.E. To make these objective values more explicit, we simultaneously listed the actual values which are corresponding to $Obj^{EA}$, $Obj^{RA}$ and $Obj^{CA}$, i.e. the total unserved energy (in kWh), the total time all the RAs take (in hours) and the total time all the CAs take (in hours), respectively. The restored loads (in kW) over time are depicted in Fig. 5. For all cases, the physical system is fully energized and all the loads (3385 kW) are restored after 263, 302, 242, and 169 minutes for Cases 1–4, respectively. To better compare the restoration processes of four cases, we present the EA's routes in Fig. 6, the RA and CA's routes in Fig. 7, as well as the switching sequences and energized node cells in Table IV.

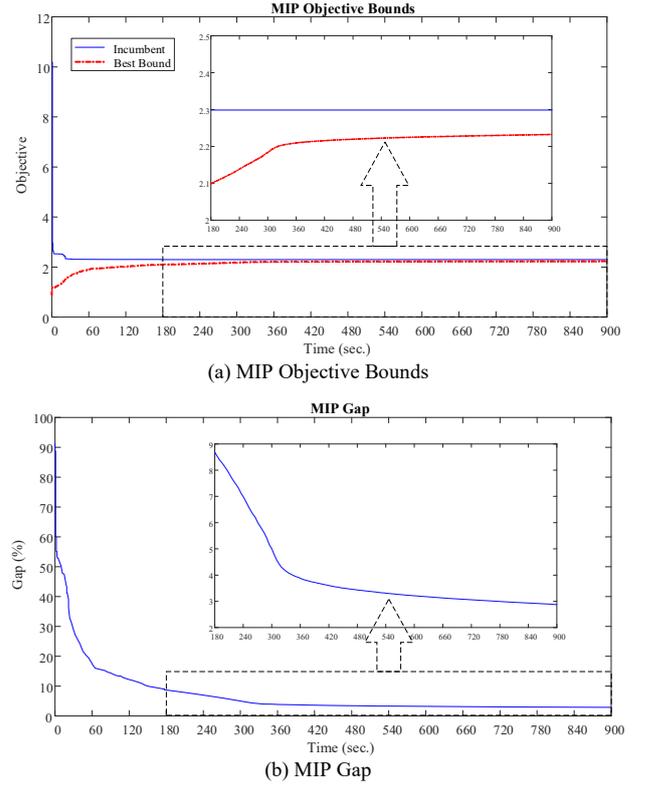

Fig. 4 Gurobi MIP solution log of Case 4

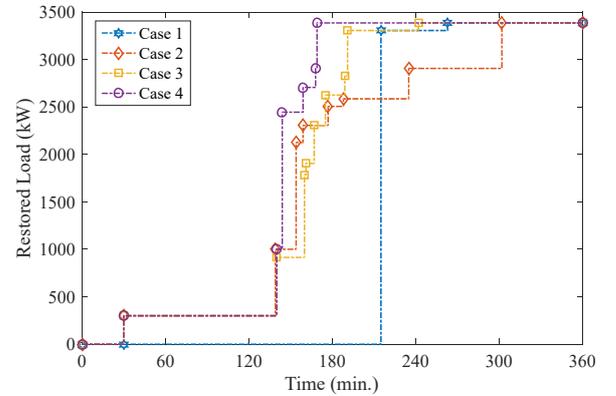

Fig. 5 Restored loads (in kW) over time in Case 1–4

TABLE III. OPTIMIZATION RESULTS OF FOUR CASES

| Case No. | Computation Time (sec.) or Gap (%) | $Obj^{EA}$ | Total Unserved Energy (kWh) | $Obj^{RA}$ | Total time all the RAs take (hours) | $Obj^{CA}$ | Total time all the CAs take (hours) | $Obj^{sum}$ |
|---|---|---|---|---|---|---|---|---|
| 1 | 1.87 sec. | 30.34% | 12193.6 | 30.56% | 14.667 | - | - | 3.3396 |
| 2 | 10.65% | 22.45% | 9646.2 | 31.32% | 15.033 | - | - | 2.5582 |
| 3 | 6.88 sec. | 22.04% | 8754.3 | 26.60% | 12.767 | 7.22% | 1.733 | 2.5422 |
| 4 | 2.88% | 19.49% | 7852.3 | 22.01% | 10.567 | 12.92% | 3.100 | 2.2983 |

As formulated in (52–53), the RA-related objective value $Obj^{RA}$ represents the average time a RA (or repair crew) takes, including travel time and work time (the latter includes repairing damaged components and manually operating switches), and the CA-related objective value $Obj^{CA}$ represents the average time a CA (or ECV) takes, including travel time and work time (i.e. duration of stay at working sites). Both objective values reflect the time cost of dispatching emergency resources to help restore power loads. The total objective values in the designed four cases are mainly decided by $Obj^{EA}$, because the weights of $Obj^{EA}$, $Obj^{RA}$, and $Obj^{CA}$ are 10, 1, and 1, which means restoring the unserved customers as quickly as possible has the highest priority compared with taking fewer emergency resources. Thus, as shown in Table III, the $Obj^{EA}$-related part takes up the most proportion of $Obj^{sum}$.

By comparing results in Table III, we can also find that: 1) Cases 3–4 have better performance than Cases 1–2 in terms of service restoration (lower $Obj^{EA}$, i.e. lower total unserved energy), which proves that the ECVs can enhance the restoration capabilities of distribution systems; 2) by comparing Case 1 and 2, we can also conclude that the crew dispatch for minimizing repair travel and work time, as indicated in the *Algorithm* 1 of Case 1, is not as good as co-optimizing the proposed "OPT-WOCA" model, in terms of service restoration. In other words, the proposed "OPT-WOCA" model can find how to enhance restoration capabilities by taking more time for RAs; 3) by comparing Cases 3 and 4, we can also conclude that dispatching ECVs by applying the "*Maximum Coverage Algorithm*" (in Case 3) is not as good as co-optimizing the proposed "OPT-WCA" model, in terms of service restoration; 4) the results of Cases 3 and 4 also highlight the necessity of considering the mobility of ECVs and the interdependencies between CAs, RAs, and EAs.

For all cases, we can compare the dynamic load restoration process in Fig. 5, which depict how much load is picked up at each time step. To observe and compare the detailed dynamic service restoration process in each case, we can check the final EA's route in Fig. 6, the RA and CA's route in Fig. 7, and the switching sequence in Table IV. In Fig. 7, the crew repair faulted lines and operate switches sequentially by traveling among these working sites, represented by the RAs' routes, which are labeled with blue arrows; the ECVs set up wireless communication networks sequentially by traveling among candidate working sites, represented by the CAs' routs, which are labeled with red arrows. As depicted in Fig. 7, all the switches in Cases 1 and Case 2 are operated manually by repair crews because there are no ECVs available to set up communication links between FTUs and the control center. Case 2 has the same EA's route as that of Case 1 but can restore unserved loads more quickly than Case 1 by solving the proposed "OPT-WOCA" model. In cases 3 and 4, as shown in Fig. 6 and Table IV, some of the automatic switches can be remotely controlled via the communication set up by ECVs at working sites. Since operating switches remotely is generally much quicker than operating them manually (e.g. 1-minute v.s. 15 minutes in our cases), the overall effect is that the restoration completion time of Case 3 and 4 is earlier than that of Case 1 and 2. In Cases 3 and 4, there are 7 and 11 automatic switches that can be operated remotely, respectively. For Case 4, More automatic switches can be operated remotely because more FTUs can be covered by ECVs at different WSs due to the movement of ECVs, and the optimal route of ECVs among WSs can be found by solving the proposed "OPT-WCA" model. The detailed operation actions of all the switches are listed in Table IV, which include the operation modes (*MD*, *ME*, *AD,* and *AE*) and operation completion time, and the sequentially energized node cells are also exhibited in detail in Table IV.

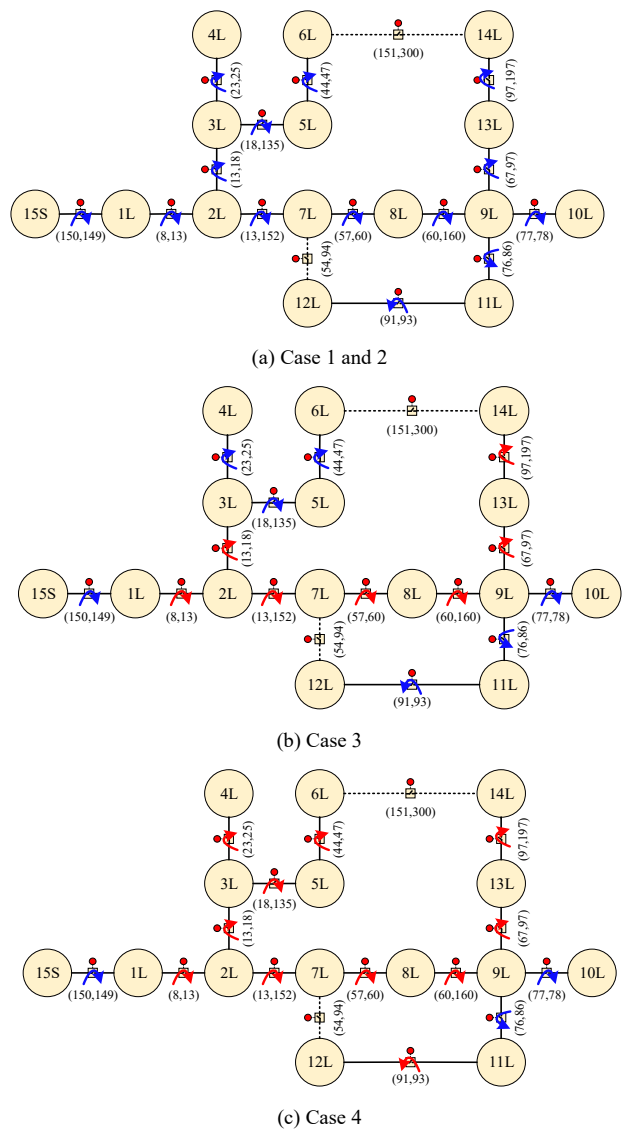

Fig. 6. The EA's route in Case 1–4.

Through the above-mentioned analysis, we can prove that

| Fig. 7 RA and CA's Route in Case 1–4 (in order) | TABLE IV. SWITCHING SEQUENCE AND ENERGIZED NODE CELLS IN CASE 1–4 (IN ORDER) |

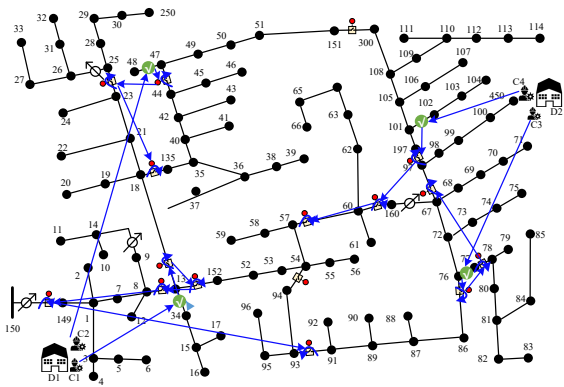

| Time (min.) | Op. (Switch) | ENCs | By Crew # | Via ECV @ WS # |
|---|---|---|---|---|
| 30 | Substation | 15S | - | - |
| 154 | MD(13,18) | - | 1 | - |
| 155 | MD(97,197) | - | 4 | - |
| 159 | MD(76,86) | - | 3 | - |
| 169 | MD(13,152) | - | 1 | - |
| 170 | MD(44,47) | - | 2 | - |
| 177 | MD(77,78) | - | 3 | - |
| 179 | MD(60,160) | - | 4 | - |
| 188 | MD(8,13) | - | 1 | - |
| 191 | MD(23,25) MD(57,60) | - | 2 4 | - |
| 201 | MD(67,97) | - | 3 | - |
| 215 | ME(150,149) MD(18,135) | 1–11L, 13L, 14L | 1 2 | - |
| 263 | ME(91,93) | 12L | 1 | - |

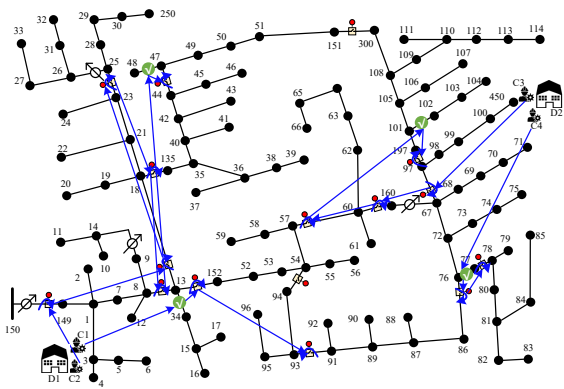

| Time (min.) | Op. (Switch) | ENCs | By Crew # | Via ECV @ WS # |
|---|---|---|---|---|
| 18 | MD(150,149) | - | 2 | - |
| 30 | Substation | 15S, 1L | - | - |
| 33 | MD(67,97) | - | 3 | - |
| 49 | MD(13,18) | - | 2 | - |
| 57 | MD(60,160) | - | 3 | - |
| 79 | MD(57,60) | - | 3 | - |
| 87 | MD(23,25) | - | 2 | - |
| 111 | MD(18,135) | - | 2 | - |
| 139 | ME(8,13) | 2-5L | 2 | - |
| 154 | ME(13,152) | 7-9L, 13L | 1 | - |
| 159 | ME(76,86) | 11L | 4 | - |
| 177 | ME(77,78) | 10L | 4 | - |
| 188 | ME(91,93) | 12L | 1 | - |
| 235 | ME(97,197) | 14L | 3 | - |
| 302 | ME(44,47) | 6L | 2 | - |

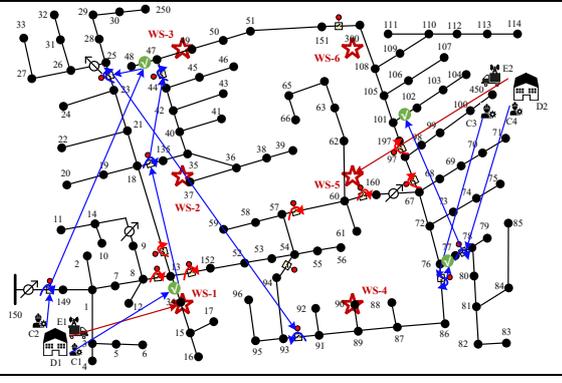

| Time (min.) | Op. (Switch) | ENCs | By Crew # | Via ECV @ WS # |
|---|---|---|---|---|
| 18 | MD(150,149) | - | 2 | - |
| 30 | Substation | 15S, 1L | - | - |
| 37 | MD(77,78) | - | 4 | - |
| 39 | MD(76,86) | - | 3 | 1 |
| 140 | AE(8,13) AD(13,152) AD(57,60) | 2L, 7L, 8L | - | 1 1 5 |
| 160 | AE(60,160) | 9-11L | - | 5 |
| 161 | AE(67,97) | 13L | - | 5 |
| 167 | AE(13,18) MD(18,135) | 3L, 5L | - 1 | 1 - |
| 175 | AE(97,197) | 14L | - | 5 |
| 189 | ME(23,25) | 4L | 2 | - |
| 191 | ME(44,47) | 6L | 1 | - |
| 242 | ME(91,93) | 12L | 2 | - |

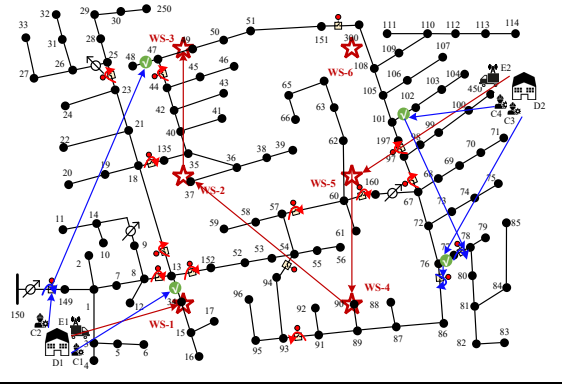

| Time (min.) | Op. (Switch) | ENCs | By Crew # | Via ECV @ WS # |
|---|---|---|---|---|
| 18 | MD(150,149) | - | 2 | - |
| 30 | Substation | 15S, 1L | - | - |
| 40 | AD(57,60) AD(60,160) AD(67,97) AD(97,197) | - | - | 5 5 5 5 |
| 70 | AD(91,93) | - | - | 4 |
| 125 | AD(18,135) | - | - | 2 |
| 140 | AE(8,13) AD(13,18) AD(23,25) | 2-5L | - | 1 1 3 |
| 144 | AE(13,152) | 7-9L, 13L, 14L | - | 1 |
| 159 | ME(76,86) | 11L | 3 | - |
| 168 | ME(77,78) | 10L | 4 | - |
| 169 | AE(44,47) | 6L | - | 3 |

**Op.(Switch)**: operation mode for switches; **MD**: manual de-energized operation; **ME**: manual energized operation; **AD**: automatic de-energized operation; **AE**: automatic energized operation; **ENCs**: energized node cells; **S**: substation cell; **L**: load cell.

the seamless coordination and cooperation of CAs, RAs, and

EAs can be fully leveraged to better enhance the restoration capabilities of distribution systems by solving the proposed "OPT-WOCA" and "OPT-WCA" models. It can also be concluded that setting up wireless communication networks by dispatching ECVs can enable the automatic switches to be controlled remotely, which reduces the travel time and the switch operation time of repair crews and speeds up the restoration of power systems. The spatial movement of the ECVs in cyber sectors causes the benefits of temporal savings in the restoration of power in physical sectors.

The 123-bus test system is medium size, of which the geographic range and the feeder's line lengths are limited. Two ECVs and 6 working sites are enough in the designed scenario considering the cover range of wireless communications set up by ECVs. The dispatch of ECVs and repair crews is essentially a vehicle routing (VR) problem [25], which is an NP-hard combinational optimization problem. If the number of working sites (i.e. the visiting targets in the VR problem), ECVs and repair crews increase, the computation complexity will increase exponentially, and solving the proposed routing model will become very complex. In this situation, we can consider some practical rules to reduce the computation complexity at the expense of optimality. For example, we can fix the CA's route table by using the proposed *Maximum Coverage Algorithm* or visiting the WSs with the maximum priority given by operators. In sum, finding an efficient, exact, and customized solution methodology to reduce the computation complexity is a challenging task, and we will study it in our future work.

## VI. CONCLUSION

In this paper, we first propose an integrated distribution system restoration framework, which considers the cooperation and coordination of the repair crews, the distribution system (physical sectors), and emergency communication networks (cyber sectors). Then, we give the specific optimization models and solution methodology of the proposed models. Finally, we conducted case studies, which validated the effectiveness and exhibited the benefit of considering ECVs and cyber-physical interdependencies in DSR. Future work includes modeling the interdependence of cyber and physical parts of distribution systems concerning situational awareness for more effective and efficient service restoration, exact and customized solution methodologies to reduce the computation complexity of the co-optimization models, etc.

ACKNOWLEDGMENT

The authors would like to thank the editor and reviewers for their insightful comments and constructive suggestions that have substantially helped to improve both the technical content and the presentation of this paper.

REFERENCES

[1] Y. Wang, C. Chen, J. Wang, and R. Baldick, "Research on Resilience of Power Systems Under Natural Disasters—A Review," *IEEE Transactions on Power Systems,* vol. 31, no. 2, pp. 1604-1613, 2016.
[2] Z. Li, M. Shahidehpour, F. Aminifar, A. Alabdulwahab, and Y. Al-Turki, "Networked Microgrids for Enhancing the Power System Resilience," *Proceedings of the IEEE,* vol. 105, no. 7, pp. 1289-1310, 2017.
[3] U. S. DOE, "Economic Benefits of Increasing Electric Grid Resilience to Weather Outages," 2013, Available: https://www.energy.gov/sites/prod/files/2013/08/f2/Grid%20Resiliency%20Report_FINAL.pdf.
[4] B. Chen, Z. Ye, C. Chen, and J. Wang, "Toward a MILP Modeling Framework for Distribution System Restoration," *IEEE Transactions on Power Systems,* vol. 34, no. 3, pp. 1749-1760, 2019.
[5] C. Li, Y. Xi, Y. Lu, and N. Liu *et al.*, "Resilient outage recovery of a distribution system: co-optimizing mobile power sources with network structure," *Protection and Control of Modern Power Systems,* vol. 7, no. 3, pp. 459-471, 2022.
[6] X. Yu and Y. Xue, "Smart Grids: A Cyber–Physical Systems Perspective," *Proceedings of the IEEE,* vol. 104, no. 5, pp. 1058-1070, 2016.
[7] B. Appasani, A. V. Jha, S. K. Mishra, and A. N. Ghazali, "Communication infrastructure for situational awareness enhancement in WAMS with optimal PMU placement," *Protection and Control of Modern Power Systems,* vol. 6, no. 1, pp. 124-135, 2021.
[8] A. Kwasinski, "Lessons from Field Damage Assessments about Communication Networks Power Supply and Infrastructure Performance during Natural Disasters with a focus on Hurricane Sandy," 2013, Available: http://users.ece.utexas.edu/~kwasinski/1569715143%20Kwasinski%20paper%20FCC-NR2013%20submitted.pdf.
[9] J. Lu, X. Xie, X. Zhou, and C. Bu, "Research on power-communication coordination recovery strategy based on grid dividing after extreme disasters," *IOP Conference Series: Earth and Environmental Science,* vol. 675, 2021.
[10] A. Arif, Z. Wang, J. Wang, and C. Chen, "Power Distribution System Outage Management With Co-Optimization of Repairs, Reconfiguration, and DG Dispatch," *IEEE Transactions on Smart Grid,* vol. 9, no. 5, pp. 4109-4118, 2018.
[11] B. Chen, Z. Ye, C. Chen, J. Wang, T. Ding, and Z. Bie, "Toward a Synthetic Model for Distribution System Restoration and Crew Dispatch," *IEEE Transactions on Power Systems,* vol. 34, no. 3, pp. 2228-2239, 2019.
[12] S. Lei, C. Chen, Y. Li, and Y. Hou, "Resilient Disaster Recovery Logistics of Distribution Systems: Co-Optimize Service Restoration With Repair Crew and Mobile Power Source Dispatch," *IEEE Transactions on Smart Grid,* vol. 10, no. 6, pp. 6187-6202, 2019.
[13] L. Xu, Q. Guo, T. Yang, and H. Sun, "Robust Routing Optimization for Smart Grids Considering Cyber-Physical Interdependence," *IEEE Transactions on Smart Grid,* vol. 10, no. 5, pp. 5620-5629, 2019.
[14] H. Lin *et al.*, "Self-Healing Attack-Resilient PMU Network for Power System Operation," *IEEE Transactions on Smart Grid,* vol. 9, no. 3, pp. 1551-1565, 2018.
[15] P. Y. Kong, "Routing in Communication Networks With Interdependent Power Grid," *IEEE/ACM Transactions on Networking,* vol. 28, no. 4, pp. 1899-1911, 2020.
[16] Y. J. Zheng, Y. C. Du, Z. L. Su, H. F. Ling, M. X. Zhang, and S. Y. Chen, "Evolutionary Human-UAV Cooperation for Transmission Network Restoration," *IEEE Transactions on Industrial Informatics,* vol. 17, no. 3, pp. 1648-1657, 2021.
[17] B. Ti, G. Li, M. Zhou, and J. Wang, "Resilience Assessment and Improvement for Cyber-Physical Power Systems Under Typhoon Disasters," *IEEE Transactions on Smart Grid,* vol. 13, no. 1, pp. 783-794, 2022.
[18] G. Huang, J. Wang, C. Chen, and C. Guo, "Cyber-Constrained Optimal Power Flow Model for Smart Grid Resilience Enhancement," *IEEE Transactions on Smart Grid,* vol. 10, no. 5, pp. 5547-5555, 2019.
[19] V. Y. Kishorbhai and N. N. Vasantbhai, "AON: A Survey on Emergency Communication Systems during a Catastrophic Disaster," *Procedia Computer Science,* vol. 115, pp. 838-845, 2017.
[20] C. Chen, J. Wang, and D. Ton, "Modernizing Distribution System Restoration to Achieve Grid Resiliency Against Extreme Weather Events: An Integrated Solution," *Proceedings of the IEEE,* vol. 105, no. 7, pp. 1267-1288, 2017.
[21] Z. Ye, C. Chen, B. Chen, and K. Wu, "Resilient Service Restoration for Unbalanced Distribution Systems With Distributed Energy Resources by Leveraging Mobile Generators," *IEEE Transactions on Industrial Informatics,* vol. 17, no. 2, pp. 1386-1396, 2021.
[22] F. X. O. Suite. (2009). *MIP formulations and linearizations - Quick reference.* Available: https://docplayer.net/53911163-Mip-formulations-and-linearizations-quick-reference.html
[23] C. IEEE PES Power System Analysis, and Economics Committee. (1992). *IEEE 123 Node Test Feeder.* Available: https://site.ieee.org/pes-testfeeders/resources/.

[24] H. Ahmadi and J. R. Martí, "Distribution System Optimization Based on a Linear Power-Flow Formulation," *IEEE Transactions on Power Delivery,* vol. 30, no. 1, pp. 25-33, 2015.
[25] G. Laporte, "Fifty Years of Vehicle Routing," *Transportation Science,* vol. 43, no. 4, pp. 407-548, 2009.

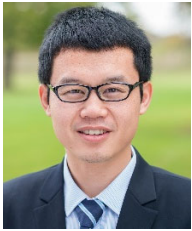

**Zhigang Ye** (Member, IEEE) received the B.S. and Ph.D. degrees from Xi'an Jiaotong University, Xi'an, China, in 2013 and 2020, respectively. He was also a visiting Ph.D. student with Texas A&M University, College Station, TX, USA, from 2016 to 2017, and with Argonne National Laboratory, Lemont, IL, USA, from 2017 to 2018.

Presently, he is a Postdoctoral Fellow with the State Grid Jiangsu Electric Power Company Ltd. Research Institute, and Southeast University, Nanjing, China. His research interests include power system reliability and resilience, operation and control of distribution systems and microgrids, demand side management, and cyber–physical systems. He is the recipient of the "2020 Top 5 Reviewers for the IEEE Transactions on Smart Grid" Award.

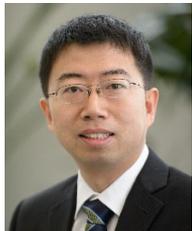

**Chen Chen** (M'13, SM'19) received the B.S. and M.S. degrees from Xi'an Jiaotong University, Xi'an, China, in 2006 and 2009, respectively, and the Ph.D. degree in electrical engineering from Lehigh University, Bethlehem, PA, USA, in 2013.

Presently, he is a Professor with the School of Electrical Engineering at Xi'an Jiaotong University, Xi'an, China. Before that, he had over six-year service at Argonne National Laboratory, Lemont, IL, USA, with the last appointment as Energy Systems Scientist at Energy Systems Division. His research interest includes power system resilience, cyber-physical system modeling and analysis, communications and signal processing for smart grid. He is the recipient of the IEEE PES Chicago Chapter Outstanding Engineer Award in 2017.

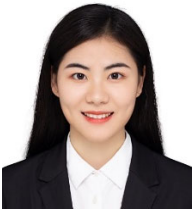

**Ruihuan Liu** (Student Member, IEEE) received the B.S. degree in electrical engineering from Xi'an Jiaotong University, Xi'an, China, in 2018. She is currently pursuing the M.S. degree in the School of Electrical Engineering, Xi'an Jiaotong University, Xi'an, China. Her major research interests include power system resilience and cyber-physical system.

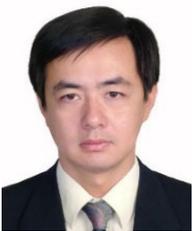

**Kai Wu** (Senior Member, IEEE) (Senior Member, IEEE) received the M.S. and Ph.D. degrees in electrical engineering from Xi'an Jiaotong University, Xi'an, China, in 1992 and 1998, respectively.

From 2000–2003, he was a Postdoctoral Fellow with Nagoya University, Nagoya, Japan. In 2003, he worked as a Research Associate with the University of Leicester, Leicester, U.K. In 2004 and 2005, he was a Visiting Researcher with the Central Research Institute of Electric Power Industry, Tokyo, Japan. Since 2006, he has been a Professor with Xi'an Jiaotong University. He is engaged in researches on electrical materials, new energy technology and integrated energy systems.

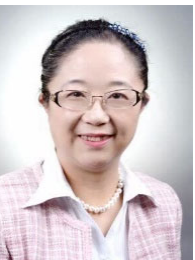

**Zhaohong Bie** (M'98, SM'12) received the B.S. and M.S. degrees from the Electric Power Department of Shandong University, Jinan, China, in 1992 and 1994, respectively, and the Ph.D. degree from Xi'an Jiaotong University (XJTU), Xi'an, China, in 1998.

Currently, she is a Professor and Vice-President of XJTU. She also serves as the Chair of IEEE Power and Energy Society China Chapter Council. Her main research interests are power system planning and reliability evaluation, integration of renewable energy, Energy Internet, microgrids, as well as resilient power systems.

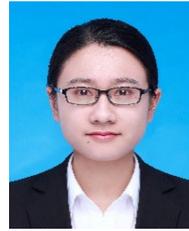

**Guannan Lou** (S'17-M'19) received the B.S. and M.S. degrees in Control Science and Engineering from North China Electric Power University, China, in 2008 and 2011, respectively. From 2011 to 2015, she joined in Guodian Nanjing Automation Co., Ltd, Nanjing. In 2018, she received the Ph.D. degree in Electrical Engineering from Southeast University, China. From 2017 to 2018, she was a joint Ph.D. student with Argonne National Laboratory.

Presently, she is an Associate Professor with the School of Electrical Engineering, Southeast University. Her research interests include distributed generations integration, networked microgrids modeling and control, resilient operation of distribution network.

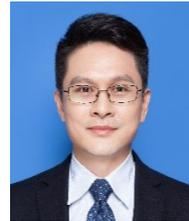

**Wei Gu** (Senior Member, IEEE) received the B.S. and Ph.D. degrees in electrical engineering from Southeast University, China, in 2001 and 2006, respectively.

From 2009 to 2010, he was a Visiting Scholar with the Department of Electrical Engineering, Arizona State University. He is currently a Professor with the School of Electrical Engineering, Southeast University. He is the Director of the Institute of Distributed Generations and Active Distribution Networks. His research interests include distributed generations and microgrids and integrated energy systems. He is an Editor for the IEEE TRANSACTIONS ON POWER SYSTEMS, the IET Energy Systems Integration, and the Automation of Electric Power Systems (China).

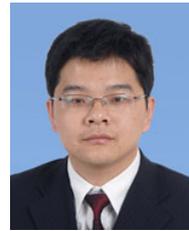

**Yubo Yuan** (Member, IEEE) received the B.S. and M.S. degrees from HoHai University, Nanjing, China, in 1997 and 2000, respectively, and the Ph.D degree from Southeast University, Nanjing, China, in 2004, all in electrical engineering.

From the last 18 years, he has been working with the State Grid Jiangsu Electric Power Company Ltd. Research Institute, where he is currently serving as the Chief Engineer. His research interests include smart distribution power grid control and optimization, control and protection of UHV DC power transmission, control and protection of flexible DC transmission, and smart substation. He is the editor for the Automation of Electric Power Systems (China). He is familiar with the work of standardization and serves as the convener of CIGRE Working Group B5/D2.67–Time in Communication Networks, Protection and Control Applications.